\documentstyle[amssymb,aps,prl,multicol]{revtex}

\newcommand{\refeq}[1]{(\ref{#1})}

\renewcommand{\paragraph}[1]{\relax}

\newif\ifcaptionsinfigures \global\captionsinfiguresfalse
\newif\ifprintfigures \global\printfiguresfalse
\newif\iffiguresintext \global\figuresintextfalse

\newif\ifmulticol \global\multicoltrue

\begin{document}
\global\firstfigfalse
\draft

\title{Critical Temperature of Bose-Einstein Condensation of 
Hard Sphere Gases}

\author{Peter Gr\"uter\thanks{\tt gruter@ncsa.uiuc.edu}
\and David Ceperley\thanks{\tt ceperley@uiuc.edu}}
\address{Department of Physics; University of Illinois;
1110 W Green Street; USA -- Urbana, Il 61801}
\author{Frank Lalo\"e\thanks{\tt laloe@physique.ens.fr}}
\address{Laboratoire Kastler Brossel; Ecole Normale Sup\'erieure;
24, rue Lhomond; F -- 75005 Paris}

\date{\today}

\maketitle

\begin{abstract}
We determine the critical temperature of a 3-d homogeneous
system of hard-sphere Bosons
by path-integral Monte Carlo simulations and
finite-size scaling.
At low densities, we find that the critical temperature is increased
by the repulsive interactions, in the form of a
power law in density with exponent $1/3$: $\Delta T_C/T_0\sim (na^3)^{1/3}$.
At high densities the result for liquid helium, namely a lower
critical temperature than in the non-interacting case, is recovered.
We give a microscopic explanation for the observed behavior.
\end{abstract}

\pacs{PACS numbers: 02.70.Lq, 03.75.Fi, 05.30.-d,-Jp}

\ifmulticol
\begin{multicols}{2}[]
\fi

\paragraph{Introduction}
The observation of Bose-Einstein condensation in atomic vapors%
\cite{atombecs},
at temperatures of a few hundreds of $nK$
by evaporative cooling, has revived interest in
the theoretical investigation of this phenomenon.
In this letter we study
bosonic hard-spheres of diameter $a$ confined in a cubic box of
volume $L^3$.
The hard-sphere diameter $a$ corresponds to the
\textit{s}-wave scattering length of a real inter-atomic
potential in a model which is valid
at low densities and even yields adequate results for
relatively dense systems such as liquid helium \cite{Kalos-1974}.

We determine the critical temperature $T_C$ of this system
for various number densities $n$ and compare it to the
critical temperature $T_0$ of the ideal gas \cite{t0formula}.
The literature provides contradictory results stemming from
analytical studies, even with regard to the sign of
$\Delta T_C=T_C-T_0$: 
a negative sign is predicted by
Hartree-Fock theory \cite{Fetter-1971-28}, and
a renormalization group calculation \cite{Toyoda-1982};
higher critical temperatures, but with different low density
asymptotic behaviors, were the result of Ref. \cite{Huang-1964},
$\Delta T_C/T_0 \sim (na^3)^{1/2}$, Ref. \cite{Stoof-1992},
$\Delta T_C/T_0 \sim (na^3)^{1/3}$, and
a recent renormalization group calculation \cite{Bijlsma-1996},
which goes beyond the one-loop expansion
of Ref. \cite{Toyoda-1982}, and which yields
$\Delta T_C/T_0\sim (na^3)^{1/6}$\cite{Stoofcomment}.

\paragraph{Path integrals and Finite-size scaling}
We have used a path-integral Monte Carlo method to obtain our results.
In 3-dimensional Bose systems both superfluidity, and Bose-Einstein
condensation, are believed to occur at the same critical temperature.
We determine this transition temperature
by making use of
the scaling properties of the superfluid fraction
$\rho_S/\rho$ \cite{Barber-1983,Pollock-1992}.
From the hypothesis that -- close to the bulk transition point --
thermodynamic quantities in finite systems
depend on temperature only through the ratio
of a characteristic system size, $L$ in our case, and
the correlation length, $\xi(t)$, 
the following scaling form can be obtained:
\begin{equation}
\rho_S/\rho(t) = L^{-\varphi/\nu} Q(L/\xi(t)),
\label{E:rhosq}
\end{equation}
valid for relative temperatures $t$ very near the critical point:
\begin{equation}
t=\frac{T-T_C}{T_C}\ll 1
\end{equation}
In Eq. \refeq{E:rhosq} $Q(L/\xi(t))$ is such a function that
--- for finite $L$ --- $\rho_S/\rho$ is
analytic in $t$ close to $t=0$.
$\varphi$ and $-\nu$ are the bulk critical exponents of the
superfluid fraction and the correlation length, respectively;
their ratio
has been measured in helium\cite{Ahlers-1973} and
calculated with renormalization group techniques\cite{LeGuillou-1977}; it
is consistent with $-\varphi/\nu=-1$.
At the transition point the correlation length diverges, so that
$Q(L/\xi(t))$ becomes independent of $L$. This feature allows us
to deduce the transition point for the infinite system
from simulations of finite size samples: curves corresponding to
various values of the scaled superfluid fraction,
$ L\rho_s/\rho(t)$, will intersect at $t=0$.

In our path-integral Monte Carlo simulation
we measure the ``stiffness'' of the system against twisting
the phase of the wave function, when a particle is displaced across the
periodic
boundary conditions, by the means of the winding number distribution.
From that we deduce the superfluid fraction \cite{pimc-superfluid}.
We use the high-temperature approximation for the
hard-sphere propagator derived in Ref. \cite{Berne-1992}.
(Our simulations converged
too slowly when the ``image propagator'' \cite{Jacucci-1983} was
employed at a first time).

For four different particle numbers, $N= 27,64,125$ and $216$,
we plot as a function of temperature the scaled quantity
\begin{equation}
N^{1/3}\rho_S/\rho(T)
\label{E:scaledrhos}
\end{equation}
in which the dimensionless $N^{1/3}$ has replaced the length $L$.

From Eq. \refeq{E:rhosq} follows that the four graphs should intersect at
one point which is to be identified as the transition point. However,
because of the statistical noise in our data, an accurate estimate
of this point is difficult (see Fig. 1).
(The same reason prevented us from
using for an estimator the scaling of the twist free energy as in
Ref. \cite{Pollock-1992}.)
According to the scaling hypothesis
$N^{1/3}\rho_S/\rho$ is analytic close to the transition point.
Therefore we
fit our values of this quantity
(for $4$ values of $N$ and $10$
different temperatures close to the transition point)
to a function of the form
\begin{equation}
\left.Q(L/\xi(T))\right|_{T\to T_C} = Q(0)+ q N^{1/(3\nu)} (T-T_C)/T_C
\end{equation}
and determine the
parameters $Q(0)$, $q$, $\nu$, and $T_C$.
In this way we obtain an quantitative estimate for $T_C$.
Furthermore we can check the results by comparing our values
for $Q(0)$ and the critical exponent $\nu$ to
results obtained elsewhere.
Assuming that the interacting
Bose gas falls into the same universality class as the XY model,
we expect that
\begin{equation}
\frac{T_0}{T_C} \frac{\zeta(\frac 32)^{2/3}}{2\pi}\ Q(0) =
\text{ universal constant}
\end{equation}
We find values between $0.29$ and $0.33$ for the
system densities we considered. This is different from the
value of $0.49\pm 0.01$ obtained in \cite{Cha-1991}.
For the critical exponent of the correlation length our
calculations yield
an estimate $-\nu=-0.68\pm 0.28$,
which is
in good agreement with analytical results \cite{LeGuillou-1980}
and simulations \cite{Li-1989} for the 3-d XY model, as well
as experimental data which give $-\nu=-0.67$.
As a matter of fact,
the values for $T_C$ only depend slightly on $\nu$: results
obtained with $\nu$ fixed to the experimental value differ at most
about $0.2\%$ from the ones obtained with letting $\nu$ be a free
parameter. 
Fig. 2 shows an exemplary fit. 

\ifprintfigures
\iffiguresintext
\begin{minipage}{\textwidth}
\begin{figure}
\input{figrs}
\end{figure}
\end{minipage}
\fi
\fi

\ifprintfigures
\iffiguresintext
\begin{minipage}{\textwidth}
\begin{figure}
\input{figrhos}
\end{figure}
\end{minipage}
\fi
\fi

\paragraph{Results}
Fig. 3 shows the results of our calculations, the
relative change of the critical temperature as a function of the density.
We distinguish two regimes:
at low densities the critical temperature for the interacting gas
is \textit{higher} than that for the non-interacting system. With
increasing density the critical temperature attains a broad
maximum (near $na^3 \approx 0.01$) before decreasing and
coming to its minimum value for the highest densities we have
used in our simulations. For comparison we include
in our diagram at higher densities experimental
and simulation results obtained
for liquid helium\cite{Wilks-1967,Pollock-1992}.
This system is well described by an hard-sphere model
with an
effective hard-sphere diameter of $a=2.2033 \text{\textit{\AA}}$
\cite{Chester-1988}.
Superfluid helium exists only for a limited range of densities before
it freezes at a density of
$na^3\approx 0.24$ \cite{Kalos-1974}.

\ifprintfigures
\iffiguresintext
\begin{minipage}{\textwidth}
\begin{figure}
\input{figtc}
\end{figure}
\end{minipage}
\fi
\fi

We have fitted the values
for densities between $na^3=5\cdot10^{-6}$ and $5\cdot10^{-3}$ against
a power law in density
\begin{equation}
\frac {\Delta T_C}{T_0}=\frac{T_C-T_0}{T_0}=c_0 (na^3)^{\gamma}
\end{equation}
and obtain $c_0=0.34\pm0.06$ and $\gamma= 0.34\pm 0.03$.
Assuming $\gamma=1/3$, the enhancement of $T_C$
is linear in the scattering length $a$.
This exponent is also the result of the analytical study
in Ref. \cite{Stoof-1992}, which yields, however, a value of
$c_0$ which is about $14$ times larger.
When mass renormalization is included, it is possible that the
RG calculation in Ref. \cite{Bijlsma-1996} may also
yield this exponent\cite{Stoofpriv}.

\paragraph{Discussion}
In order to understand our findings on a microscopic level,
recall the Feynman picture of superfluidity
\cite{Feynmansuperfl}.
The partition function $\mathcal{Z}$ of the system can
be written as a sum over all states accessible to
distinguishable particles:
\begin{equation}
{\mathcal Z}=\mathop{\text{Trace}}_{\text{Boltzmann states}}
\left\{ S\ e^{-\beta
{\mathcal H}}\right\}
\end{equation}
where $S$ is the symetrizer with respect to the particle labels.
We develop the symetrizer $S$ into a sum over exchange
cycles.
At high temperatures the only significant contribution
comes from the identity operator (cycle length $1$) and
$\mathcal Z$ reduces to the partition function
for a classical system.
At lower and lower temperatures, however,
the importance of the contribution of exchange
cycles of macroscopic length grows.
The statistical correlations extend over longer and
longer distances, until -- at the transition
point -- a macroscopic structure, identified
as superfluid component, is established.

A necessary condition for these exchange cycles to appear is that
the particles must be close together
(separated by distances of the order of the
thermal wavelength $\lambda_T=2\pi\hbar^2/(m k_B T)$).
In an ideal gas, spatial density fluctuations are important, and
particle clusters are likely to appear. This creates
regions of lower density at other places in the system, which
obstruct the formation of macroscopic exchange cycles.
On the other hand, in the case of an moderately
dense interacting gas, the particles tend to be more homogeneously
distributed throughout the whole volume of the system.
Analogously to percolation problems, it is in this case more
likely for every atom to find a neighbor at a
suitable distance, so that the exchange cycles can more easily
propagate from one particle to the next. Superfluidity is
hence ``easier'' to achieve, meaning that it can occur at higher
temperatures.

\ifprintfigures
\iffiguresintext
\begin{minipage}{\textwidth}
\begin{figure}
\input{figgr}
\end{figure}
\end{minipage}
\fi
\fi

In order to quantify this understanding
of the shift of the critical temperature, we have determined the
two-body distribution function $g(r)$. In Fig. 4
we have plotted $g(r)$ for different densities at
the bulk superfluid transition temperature for a system of $125$ particles.
It can be seen easily that at higher densities the ``exchange
bump'' of the ideal gas decreases and, because of
the hard-sphere boundary condition, the distribution tends to become
homogeneous. Another way to see that the interactions suppress
the spatial density fluctuations is to integrate $g(r)$ over
sub-volumes of the simulation cell
which gives the fluctuation of the particle number in
these sub-volumes\cite{Pathria143}.
We find that the number of particles in
sub-spheres of radius comparable to $\lambda_T$ fluctuates
about $30\%$ more in the case of an ideal gas than for
an interacting gas of density $5\cdot 10^{-2} a^{-3}$. 

At higher densities the atoms cannot exchange with each other
without dragging other particles along. The
effective mass becomes greater than unity, and the critical temperature is
decreased as $T_C\sim 1/m^{\star}$ \cite{Fetter-1971-28,Feynmansuperfl}.

The results obtained in our case of an isotropic system are
different from the ones obtained in an harmonic external potential
\cite{Giorgini-1996,Krauth-1996}, where the repulsive interactions decrease the
critical temperature at low densities.
In a trap, interactions give rise to two competing effects. The critical
temperature is increased by the suppression of density fluctuations
(even if those are already smaller than in the isotropic case because of the
confinement). On the other hand, the central density is lowered as the
condensate wave function is broadened. Both terms are linear in the scattering
length; hence, whether $T_C$ is suppressed or increased depends on
details of the system such as the number of particles, the trapping
frequency $\omega$ and the particle mass\cite{Stooftrap}.

To conclude, the effect of
repulsive interactions is different for
low and high densities:
at low densities, the ``density homogenization'' effect prevails
so that $T_C$ is increased; at high densities, the exchange motion
of the atoms -- which are behind the Bose-Einstein condensation
phenomenon - are hindered by the hard cores so that the critical
temperature is decreased. Since, in the experiments on alkali gases,
the value of $na^3$ in the center of the trap does not exceed
$10^{-5}$, the shifts in $T_C$ remain small (about $1\%$), but
may be observable in larger traps.

The authors thank M. Dewing and W. Krauth for helpful comments and discussions.
This project was supported by QNR-N\textbf{00014}-90-J-1783. The computations
were performed at the National Center for Supercomputing Applications,
Champaign-Urbana, Illinois. The Laboratoire Kastler Brossel is
``laboratoire associ\'e au CNRS et \`a l'Universit\'e Paris VI''.
\newcommand{\noopsort}[1]{} \newcommand{\printfirst}[2]{#1}
  \newcommand{\singleletter}[1]{#1} \newcommand{\switchargs}[2]{#2#1}
  \providecommand{\bibchapter}{chapter }\providecommand{\bibtobepublished}{to
  be published}\providecommand{\bibsecond}{Second}

\ifcaptionsinfigures
\else
\clearpage
\textbf{Figure captions}

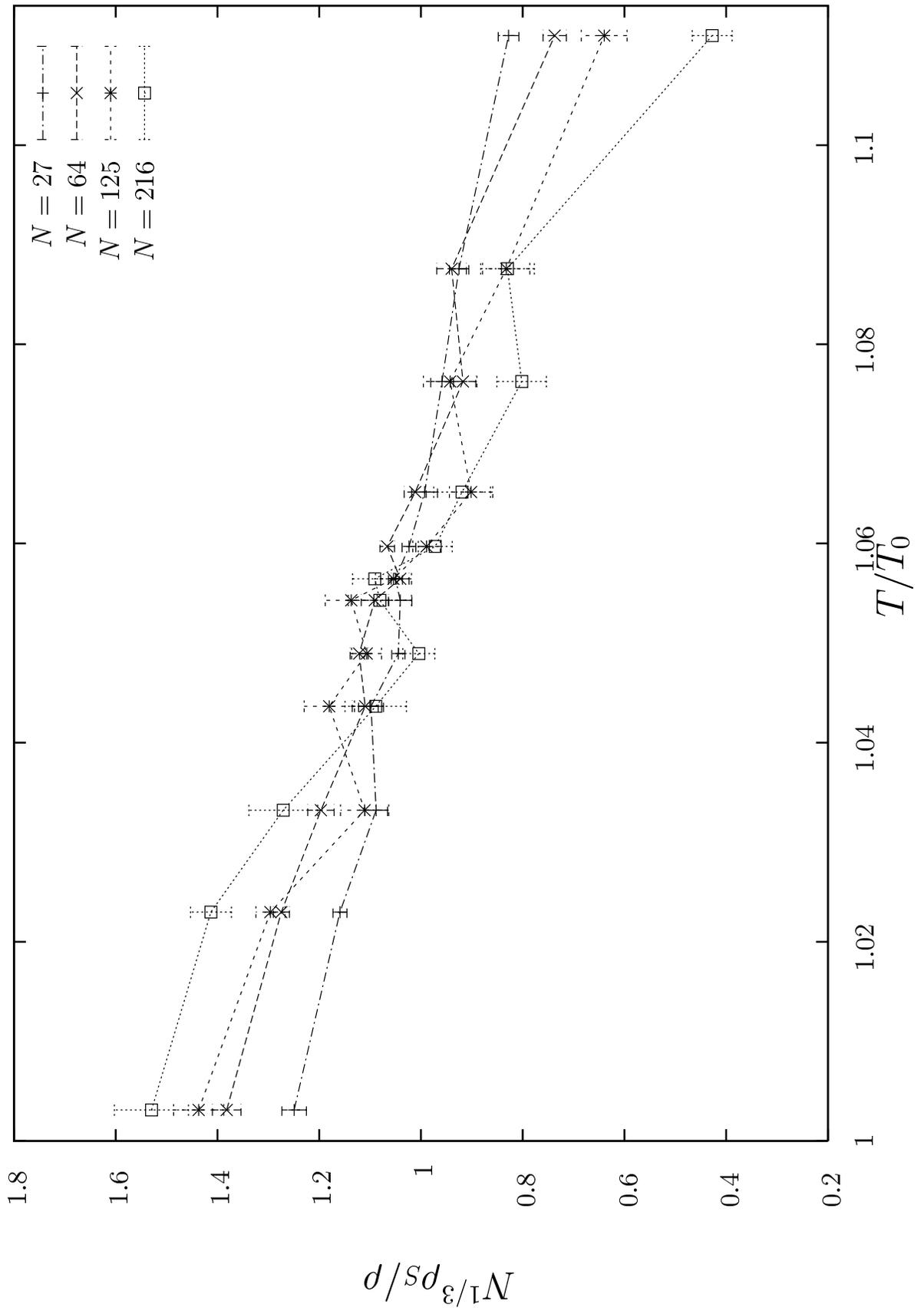
\noindent\underline{Figure 1:}\\
Scaled superfluid fraction
$N^{1/3}\rho_S/\rho$ in function of temperature $T$
as determined by our simulations for
number density $na^3=5\cdot 10^{-3}$ and
$N=27$,$64$, $125$, and $216$. $T_0$ is the critical
temperature of the non-interacting system. The temperature
where the four lines are crossing is to be identified with the
critical temperature. A quantitative estimate for this
temperature is obtained by fitting the values for every $N$
with a straight line and determining the point of intersection for
the fitted curves.

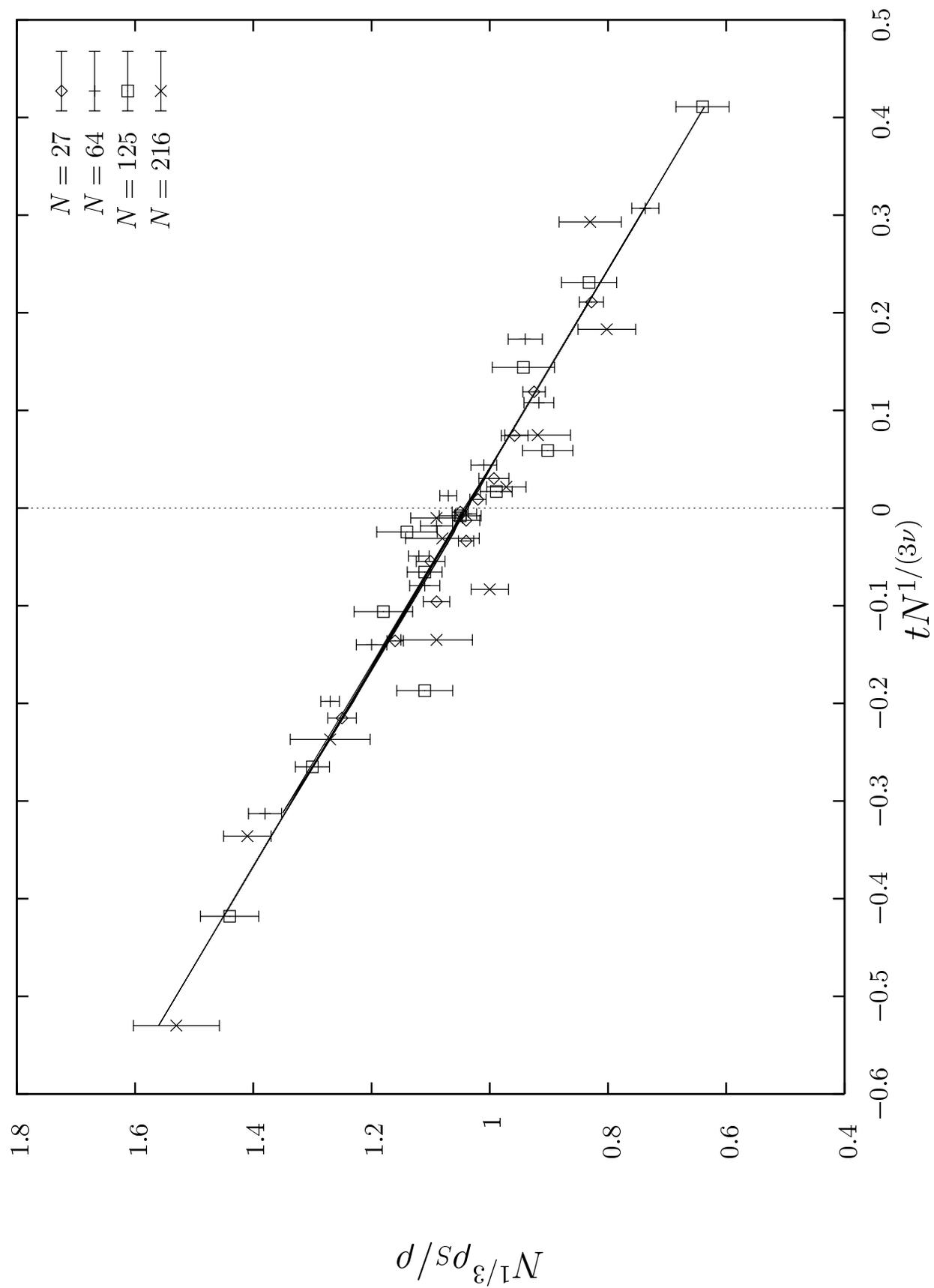
\noindent\underline{Figure 2:}\\
Result of our fit to the data from
Fig. 1.  $t=(T-T_C)/T_C$,
$T_C$ is the critical temperature for
the interacting system,  $-\nu$ 
the critical exponent of the correlation length, both as obtained
by our fit. We find
a critical temperature of $T_C/T_0=1.057 \pm 0.002$

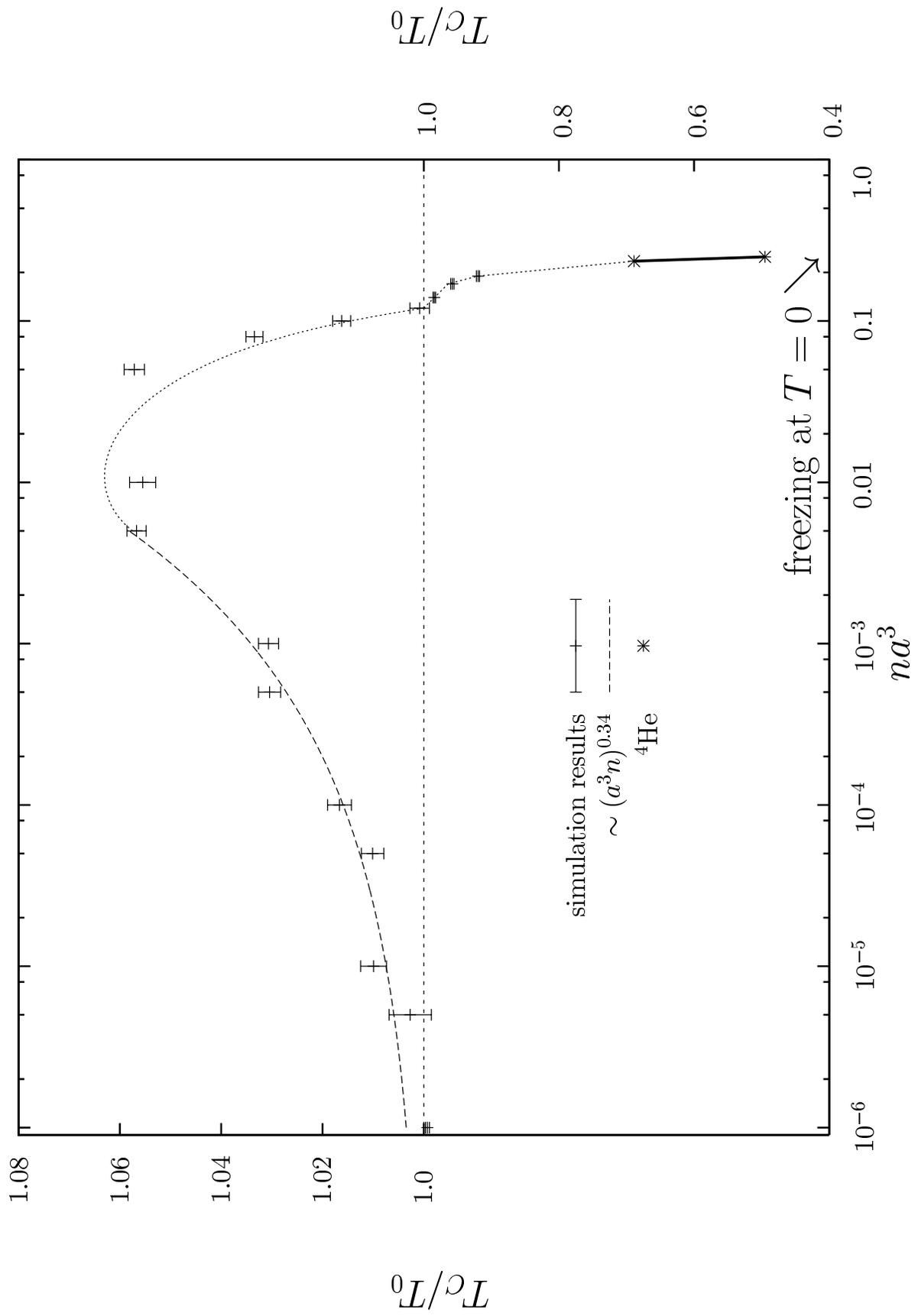
\noindent\underline{Figure 3:}\\
Critical temperature $T_C$ of an interacting Bose gas versus
density: $a$ is the hard-sphere diameter, $T_0$ the critical
temperature of the non-interacting gas.
Two different scales are used for the vertical axis:
for temperatures above $T_0$ the left scale,
for values below $T_0$ the right scale apply.
For comparison, experimental and simulation results
\protect\cite{Wilks-1967,Pollock-1992}
obtained for helium (at densities $na^3=0.235$ and $0.25$)
are also included.
The dashed curve at low densities presents a fit
of the data points between $na^3=5\cdot 10^{-6}$ and
$5\cdot 10^{-3}$ to $1+c_0 (na^3)^{\gamma}$,
yielding $\gamma=0.34$, the dotted curve is guide to the eye
(note the change of scale at $T_C=T_0$).
At zero temperature the hard-sphere system freezes at a density of
about $na^3=0.25$\protect\cite{Kalos-1974}.

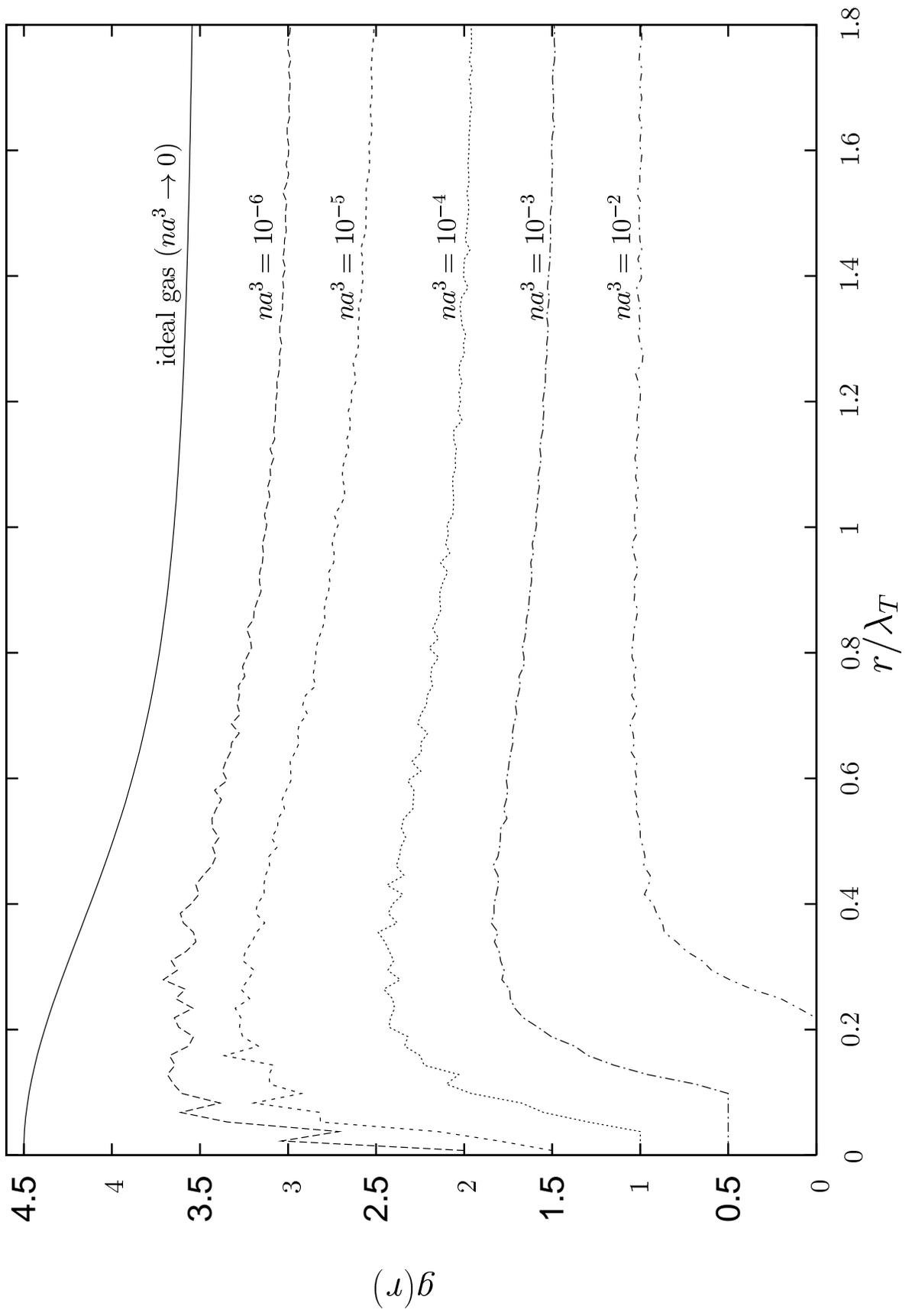
\noindent\underline{Figure 4:}\\
Two-body distribution function $g(r)$ versus particle
distance $r$ in units of the thermal wavelength $\lambda_T$ for
different densities $na^3$ at the bulk critical temperature.
For the sake of visibility the single curves are shifted vertically
by an amout of $0.5$.
The hard-sphere boundary condition
changes the short-distance ($r\to0$) behaviour dramatically.
However, for the ``percolation''-like theory of superfluidity the
behaviour for $r\approx\lambda_T$ is more important, and here the
functions for the ideal and the low density gas both feature the
appearance of the ``exchange bump'' for $r\lesssim\lambda_T$:
in the non-interacting gas or
gases of lower density,
many particles cluster together at distances less than
the thermal wavelength. In denser system the interactions
are compensating this statistical attraction and the particles
fill the system volume homogeneously. This allows for superfluidity
to appear at higher temperatures.
\fi

% figures
\ifprintfigures
\iffiguresintext
\else
\clearpage
\begin{figure}
\input{figrs}
\end{figure}
\clearpage
\begin{figure}
\input{figrhos}
\end{figure}

\clearpage
\begin{figure}
\input{figtc}
\end{figure}

\clearpage
\begin{figure}
\input{figgr}
\end{figure}

\fi
\fi

\ifmulticol
\end{multicols}
\fi


\begin{thebibliography}{10}
\bibitem{t0formula}
$T_0=(\hbar^2 2\pi/mk_B)[n/\zeta(\frac 32)]^{2/3}$ where $m$ is the particle
  mass, $k_B$ the Boltzmann constant, and $\zeta(\frac 32)\approx2.612$.
\end{thebibliography}
\end{document}